%% file: spec.tex
\documentclass[twoside]{sfm}

\input{sf.def}

\begin{document}

\def\llm{{\sc LLmodels}}
\def\atl{{\sc ATLAS9}}
\def\aatl{{\sc ATLAS12}}
\def\starsp{{\sc STARSP}}
\def\aur{$\Theta$~Aur}
\def\logg{\log g}
\def\tauros{\tau_{\rm Ross}}
\def\kms{km\,s$^{-1}$}
\def\bz{$\langle B_{\rm z} \rangle$}
\def\degr{^\circ}
\def\aaps{A\&AS}
\def\aap{A\&A}
\def\apjs{ApJS}
\def\apj{ApJ}
\def\rmxaa{Rev. Mexicana Astron. Astrofis.}
\def\mnras{MNRAS}
\def\actaa{Acta Astron.}
\newcommand{\Tef}{T$_{\rm eff}$~}
\newcommand{\Vt}{$V_t$}
\newcommand{\CC}{$^{12}$C/$^{13}$C~}
\newcommand{\CDC}{$^{12}$C/$^{13}$C~}

\input{schoeller/schoeller.tex}

\end{document}

%% file: schoeller/schoeller.tex
\pagebreak

\thispagestyle{titlehead}

\setcounter{section}{0}
\setcounter{figure}{0}
\setcounter{table}{0}

\markboth{Sch\"oller et al.}{$\beta$\,Cep, SPB, and Be stars}

\titl{Magnetic fields in $\beta$\,Cep, SPB, and Be stars}{Sch\"oller M.$^1$, Hubrig S.$^2$, Briquet M.$^3$, Ilyin I.$^2$}
{$^1$European Southern Observatory, Karl-Schwarzschild-Str.~2, 85748~Garching, Germany, email: {\tt mschoell@eso.org} \\
 $^2$Leibniz-Institut f\"ur Astrophysik, An~der Sternwarte 16, 14482~Potsdam, Germany\\
 $^3$Institut d'Astrophysique et de G\'eophysique, Universit\'e de Li\`ege, All\'ee du 6~Ao\^ut~17, B\^at~B5c, 4000~Li\`ege, Belgium}

\abstre{
Recent observational and theoretical results emphasize the potential
significance of magnetic fields for structure, evolution, and environment of
massive stars.
Depending on their spectral and photometric behavior, the upper
main-sequence B-type stars are assigned to different groups, such as $\beta$\,Cep
stars and slowly pulsating B (SPB) stars, He-rich and He-deficient Bp stars,
Be stars, BpSi stars, HgMn stars, or normal B-type stars.
All these groups are
characterized by different magnetic field geometry and strength, from fields
below the detection limit of a few Gauss up to tens of kG.
Our collaboration was the first to systematically study the magnetic fields
in representative samples of different types of main-sequence B stars.
In this article, we
give an overview about what we have learned during the last years about magnetic
fields in $\beta$\,Cep, SPB, and Be stars.
}

\baselineskip 12pt

\section{Magnetic fields in massive stars}
\label{sect:schoeller_intro}

The presence of a convective envelope is a necessary condition
for significant magnetic activity.
Magnetic activity is found all the way from the late A-type stars
(e.g.\ in Altair: Robrade \& Schmitt 2009 \cite{schoeller_RobradeSchmitt2009})
with very shallow convective envelopes down to the coolest fully convective M-type stars.

On the other hand, advances in instrumentation over the past decades
have led to magnetic field detections in a small
but gradually growing subset of massive stars,
which frequently present cyclic wind variability,
H$\alpha$ emission variations, non-thermal radio/X-ray emission,
 and transient features in absorption line profiles. 

Magnetic fields have fundamental effects on the evolution of massive stars, 
their rotation, and on the structure, dynamics, and heating of radiative winds.
During the last years, an increasing number of massive stars have been investigated for magnetic fields. 
Currently, more than two dozen magnetic early B-type stars
(excluding classical He-strong/He-weak Bp stars) are known. 

The origin of the magnetic fields is still under debate:
it has been argued that magnetic fields could be ``fossil'',
or magnetic fields may be generated by strong binary interaction,
i.e.\ in stellar mergers, or during a mass transfer or common envelope evolution.

\section{Determining magnetic fields with FORS\,1/2}
\label{sect:schoeller_determ}


\begin{figure}[!t]
\begin{center}
 \includegraphics[width=0.45\textwidth]{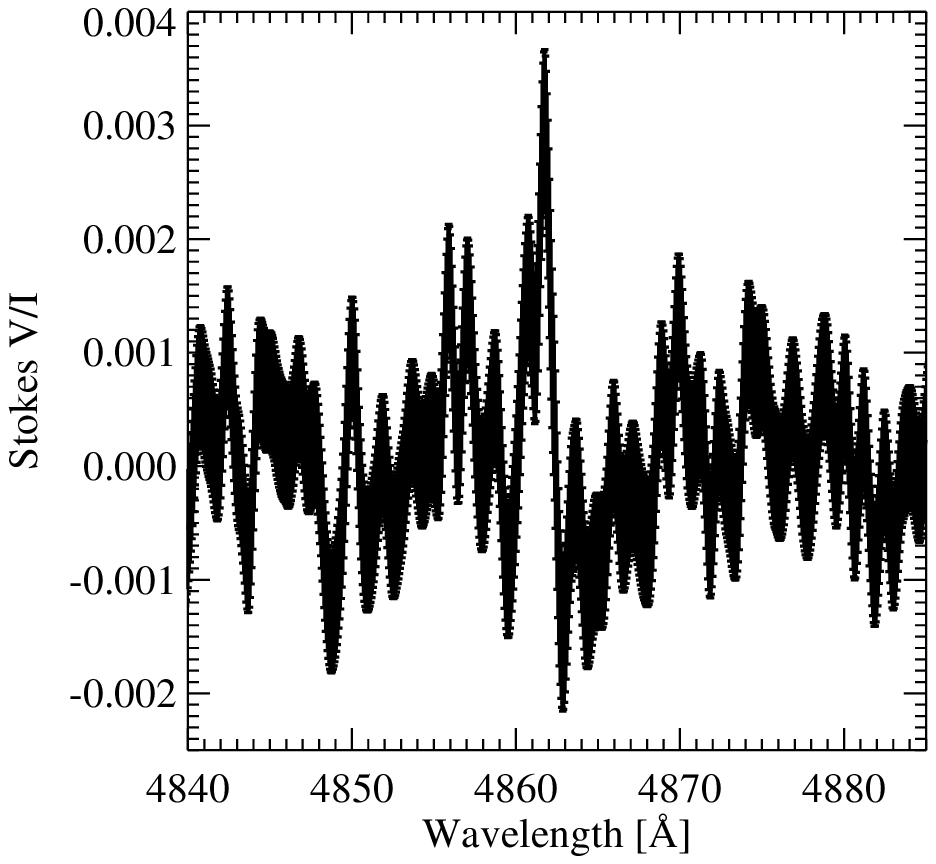}
 \includegraphics[width=0.45\textwidth]{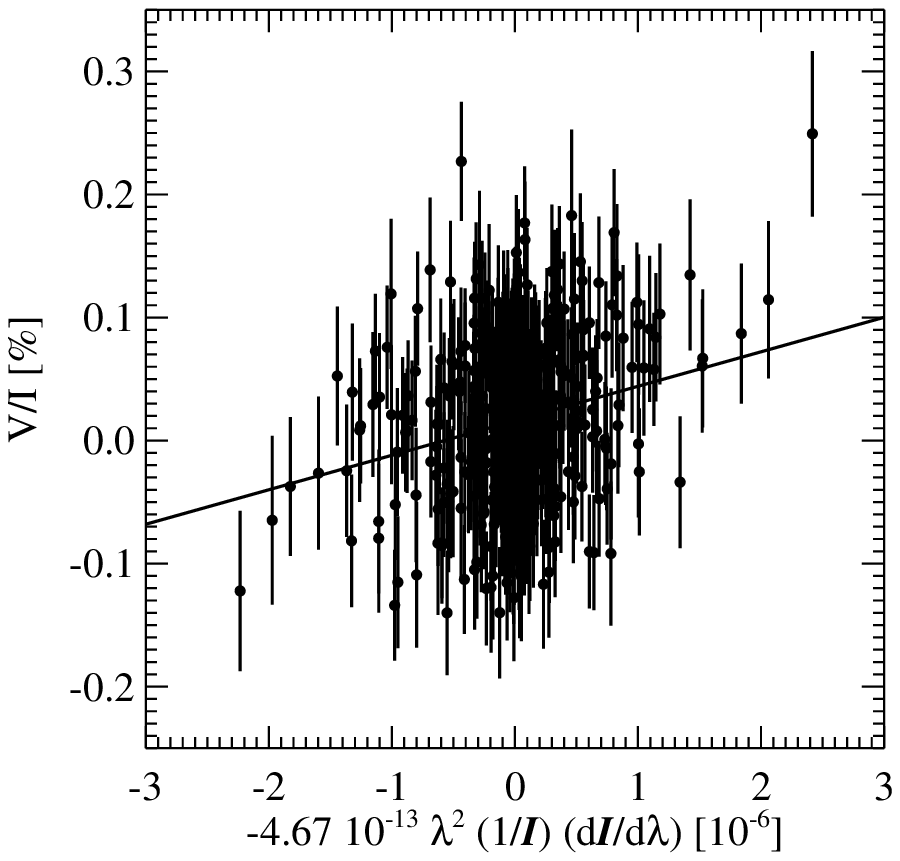}
\vspace{-5mm}
\caption[]{
{\em Left:} Part of the Stokes~$V$ spectrum of $\xi^1$\,CMa around H$\beta$ with a typical Zeeman pattern.
{\em Right:} Regression detection of a magnetic field in $\xi^1$\,CMa.
}
\label{fig:schoeller_VandRegression}
\end{center}
\end{figure}

FORS\,2 is a multi-mode instrument equipped with
polarization analyzing optics comprising super-achromatic half-wave and quarter-wave
phase retarder plates, and a Wollaston prism with a beam divergence of 22$^{\prime\prime}$  in
standard resolution mode.
Before the polarimetric optics was moved to FORS\,2, it was installed in its twin, FORS\,1.
From the FORS\,2 data, the Stokes~$V/I$ spectrum is calculated following:

\begin{equation}
\frac{V}{I} = \frac{1}{2} \left( \left( \frac{f^o - f^e}{f^o + f^e}\right)_{\alpha = -45^{\circ}} -
\left( \frac{f^o - f^e}{f^o + f^e}\right)_{\alpha = +45^{\circ}} \right)
\end{equation}

\noindent
where $\alpha$ gives the position angle of the retarder waveplate
and $f^o$ and $f^e$ are the ordinary and extraordinary beams, respectively.

The mean longitudinal magnetic field is the
component of the magnetic field parallel to the line of sight,
averaged over the stellar hemisphere visible at the time of observation.
It is diagnosed from the slope of the linear regression:

\begin{equation}
\frac{V}{I} = -\frac{ g_{\rm eff}\,e}{4\pi{}m_ec^2} \lambda^2 \frac{1}{I} \frac{{\rm d} I}{\rm{d} \lambda} \langle B_z \rangle
\end{equation}

\noindent
where
$V$ is the Stokes parameter that measures the circular polarization,
$I$ is the intensity in the unpolarized spectrum,
$g_{\rm eff}$ is the effective Land\'e factor,
$e$ is the electron charge,
$\lambda$ is the wavelength,
$m_e$ the electron mass,
$c$ the speed of light,
${{\rm d}I/{\rm d}\lambda}$ is the derivative of Stokes $I$,
and $\left<B_{\rm z}\right>$ is the mean longitudinal magnetic field.
A typical regression detection can be found in Fig.~\ref{fig:schoeller_VandRegression}.

\section{Magnetic fields in B-type stars}
\label{sect:schoeller_mf_b}

Depending on their spectral and photometric behavior,
the main-sequence B-type stars are assigned to different groups,
such as $\beta$\,Cep stars and slowly pulsating B (SPB) stars,
He-rich and He-deficient Bp stars,
Be stars, BpSi stars, HgMn stars, or normal B-type stars. 
These groups are characterized by different magnetic field geometry and strength,
from fields below the detection limit of a few Gauss up to tens of kG. 

To identify and to model the physical processes responsible for the generation
of magnetic fields in massive stars, it is important to understand whether:
\begin{itemize}
\item most magnetic stars are slowly rotating
\item magnetic fields appear in stars at a certain age
\item magnetic fields are generated in stars in special environments:
Do some clusters contain a larger number of magnetic massive stars,
similar to the Ap/Bp content in different clusters
(NGC 2516 has the largest number of magnetic Ap stars and X-ray sources)?
\item magnetic fields are produced through binary interaction
\item X-ray emission can be used as an indirect indicator for the presence of magnetic fields
\end{itemize}

\subsection{Pulsating B stars}
\label{subsect:schoeller_mf_pulsb}

$\beta$\,Cep stars are short-period (3--8\,h) pulsating variables of
spectral type O9 to B3 (corresponding to a mass range of 8--20\,M$_\odot$)
along the main sequence that pulsate in low-order pressure (p) and/or gravity (g) modes. 
SPB stars show variability with  periods of the order of 1\,d, are
less massive (3--9\,M$_\odot$) main sequence B-type stars and have
multiperiodic high-order low-degree g mode oscillations.

\begin{figure}[!t]
\begin{center}
 \includegraphics[width=0.7\textwidth]{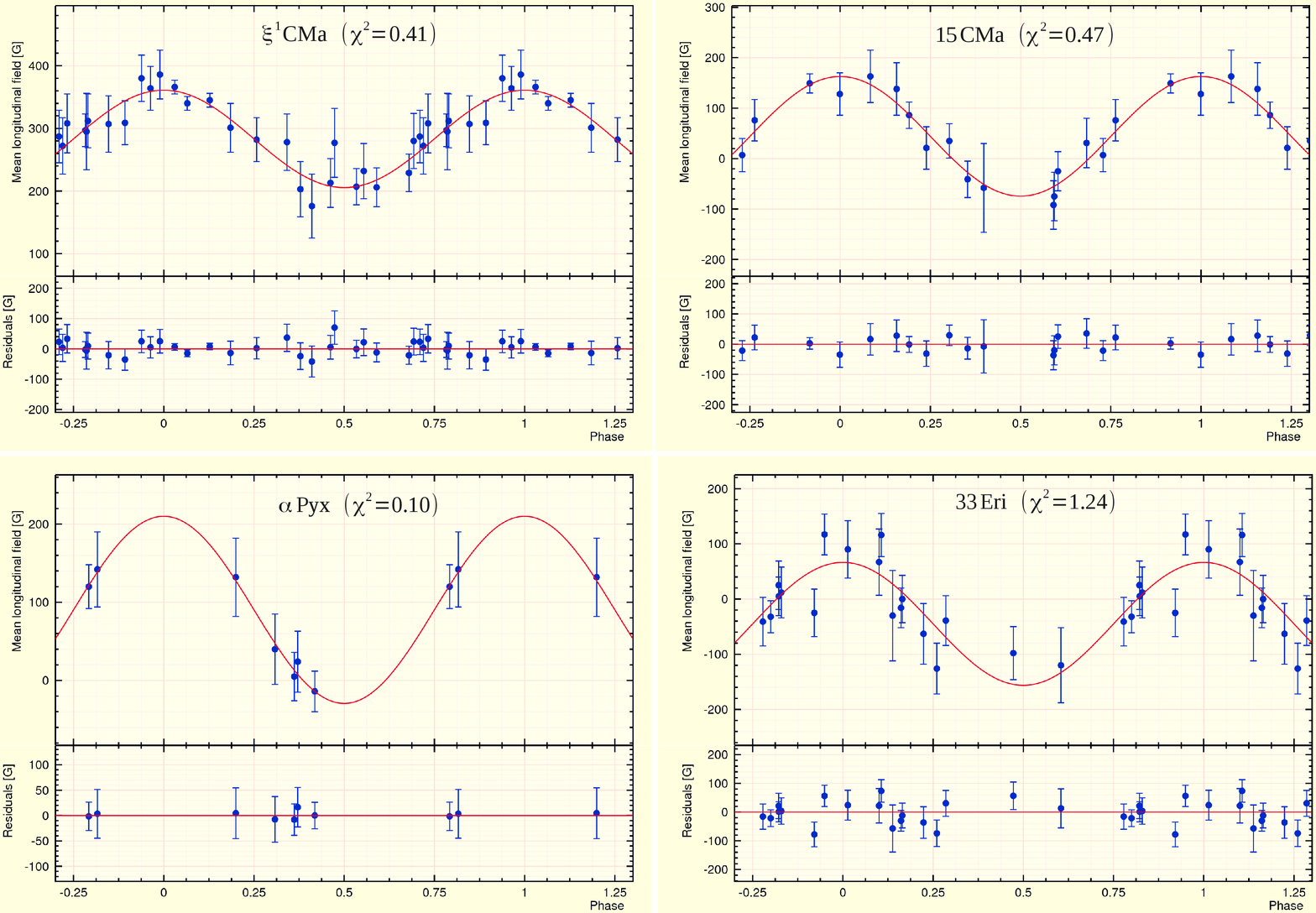}
\vspace{-5mm}
\caption[]{
Phase diagrams with the best sinusoidal fit for the longitudinal magnetic field measurements.
The residuals (Observed -- Calculated) are shown in the lower panels. The deviations are mostly of the same
order as the error bars, and no systematic trends are obvious, which justifies a single sinusoid as a
fit function.
}
\label{fig:schoeller_phased_mfs}
\end{center}
\end{figure}

A long-term monitoring project aimed at asteroseismology
of a large sample of slowly pulsating B (SPB) stars
and $\beta$\,Cep stars was started by researchers
of the Institute of Astronomy of the University of Leuven more than ten years ago. 
In our first publication on a magnetic survey of pulsating B-type stars with FORS\,1
(Hubrig et al.\ 2006 \cite{schoeller_Hubrig2006}),
we announced detections of a weak mean longitudinal magnetic field of the order
of a few hundred Gauss in a number of SPB stars
and in the $\beta$\,Cep star $\xi^1$\,CMa,
whose field, of the order of 300--400\,G, is one of the largest among all currently
known magnetic $\beta$\,Cep stars.
In Fig.~\ref{fig:schoeller_phased_mfs}, we display our results of magnetic field
monitoring of four $\beta$\,Cep and SPB stars (Hubrig et al.\ 2011a \cite{schoeller_Hubrig2011a}).
From FORS\,1/2 and SOFIN observations, we determined a rotation period of
$P = 2.1795$\,d for $\xi^1$\,CMa,
which is not in line with
Fourtune-Ravard et al.\ (2011 \cite{schoeller_FourtuneRavard2011}), who determined $P \sim 4.2680$\,d 
from ESPaDOnS observations.
Note that in that work, the impact of pulsations on the magnetic field
measurements from high resolution spectra was not taken into account.

\begin{table}
\centering
\caption{Measurements of the mean longitudinal magnetic field using high-resolution HARPS spectra.}
\label{tab:schoeller_mf_HARPS}
\tabcolsep1.2mm
\begin{tabular}{lrrr@{$\pm$}r}
\hline
\hline
\multicolumn{1}{c}{Object} &
\multicolumn{1}{c}{MJD} &
\multicolumn{1}{c}{S/N} &
\multicolumn{2}{c}{$\left<B_{\rm z}\right>$} \\
\hline
HD\,74195 & 55605.217 & 220 & $-$70 & 21 \\
HD\,74195 & 55606.130 & 300 & $-$14 & 18 \\
HD\,74560 & 55605.221 & 240 &    56 & 19 \\
HD\,74560 & 55606.134 & 280 &     8 & 18 \\
HD\,74560 & 55607.177 & 350 & $-$35 & 15 \\
HD\,85953 & 55600.305 & 230 &    79 & 20 \\
\hline
\end{tabular}
\end{table}

Among the sample of SPB stars with detected magnetic fields using FORS\,1,
three stars, HD\,74195, HD\,74560, and HD\,85953, have been observed
in 2011 February with the high-resolution ($R = 115,000$) polarimeter HARPSpol,
installed at the ESO 3.6\,m telescope on La Silla,
in the framework of the GTO program 086.D-0240(A).
The star HD\,85953 was observed once,
whereas HD\,74195 was observed on two different nights,
and HD\,74560 was observed on three different nights.
We downloaded from the ESO archive the available spectra
and reduced them using the HARPS data reduction software
available at the ESO headquarters in Germany.
For the measurements of the magnetic fields,
we used the moment technique developed by Mathys (e.g.\ Mathys 1991 \cite{schoeller_Mathys1991}).
Formally significant detections above the 3$\sigma$ level were achieved in
HD\,85953 and in one observation of HD\,74195 (see Table~\ref{tab:schoeller_mf_HARPS};
Hubrig et al.\ 2013, {\em in preparation}).
In line with our discoveries of the presence of weak magnetic fields in pulsating stars,
Briquet et al.\ (2013 \cite{schoeller_Briquet2013}) found a magnetic field
in the hybrid SPB/$\beta$\,Cep star HD\,43317.


\begin{figure}[!t]
\begin{center}
\includegraphics[width=0.32\textwidth, angle=0]{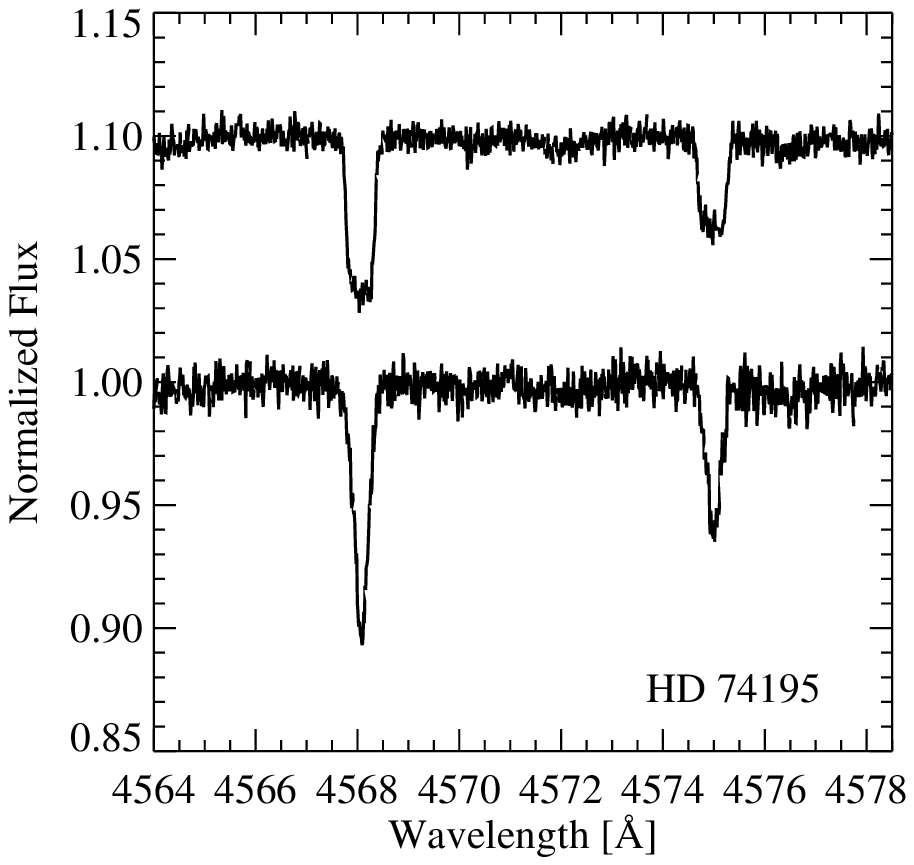}
\includegraphics[width=0.32\textwidth, angle=0]{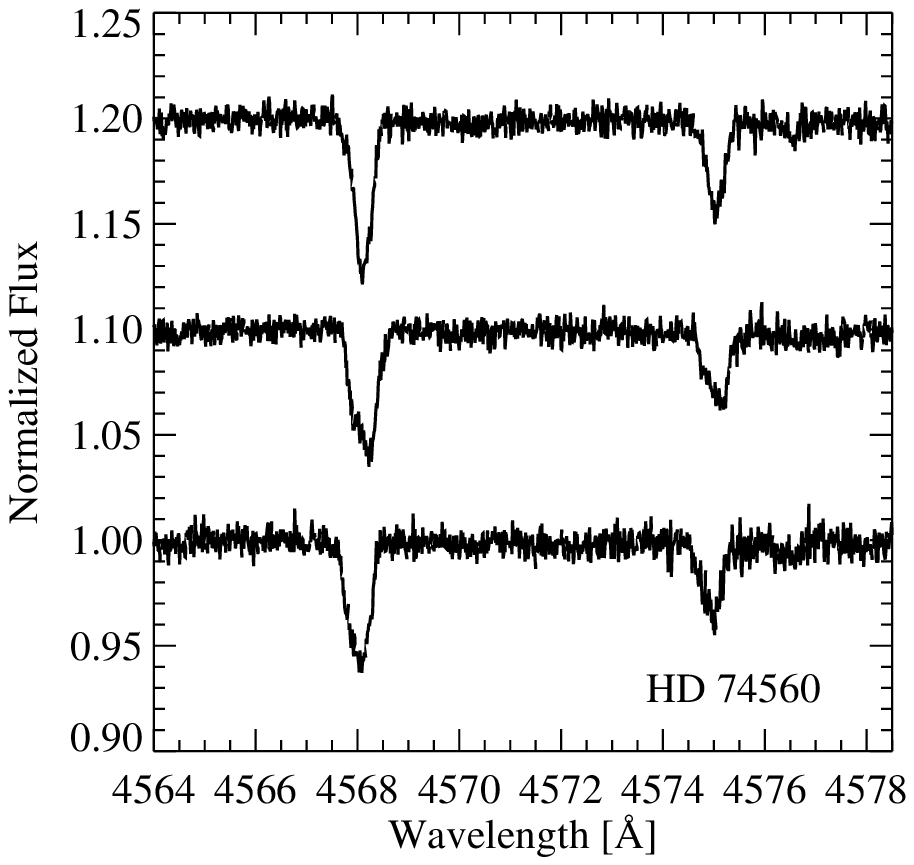}
\includegraphics[width=0.32\textwidth, angle=0]{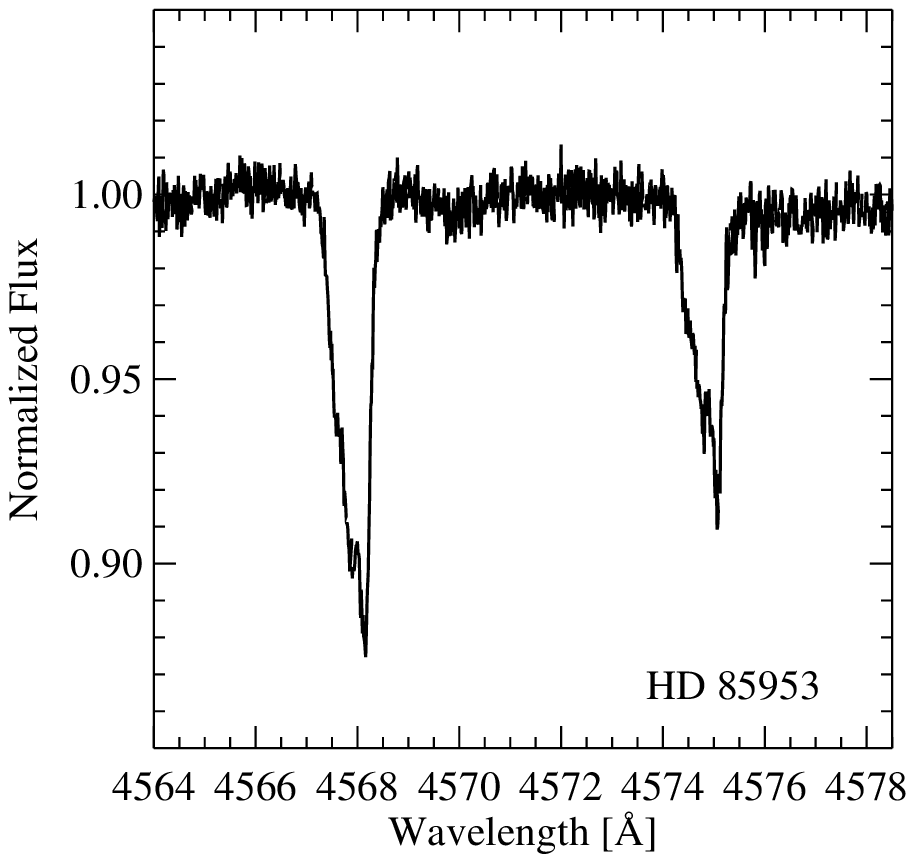}
\includegraphics[width=0.32\textwidth, angle=0]{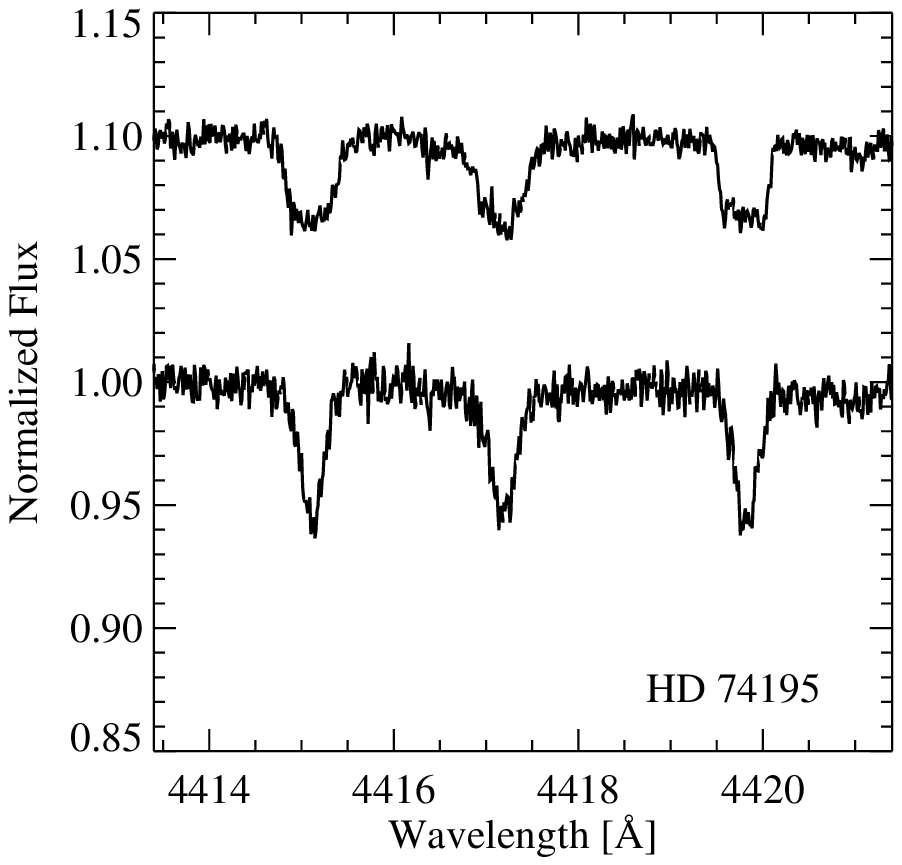}
\includegraphics[width=0.32\textwidth, angle=0]{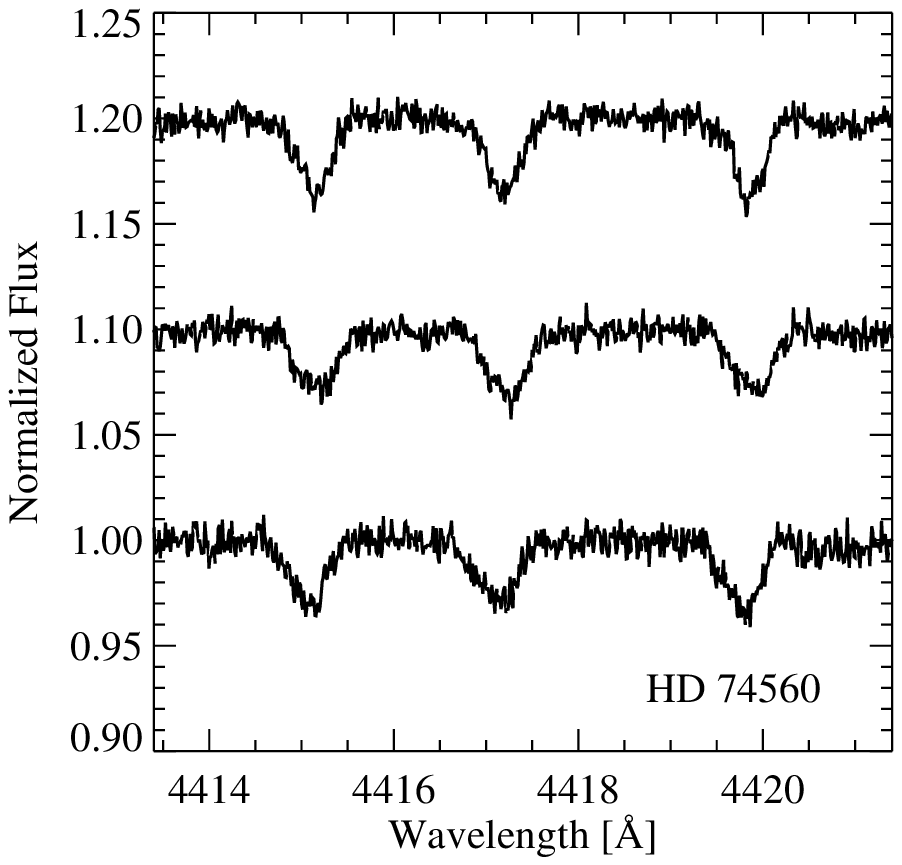}
\includegraphics[width=0.32\textwidth, angle=0]{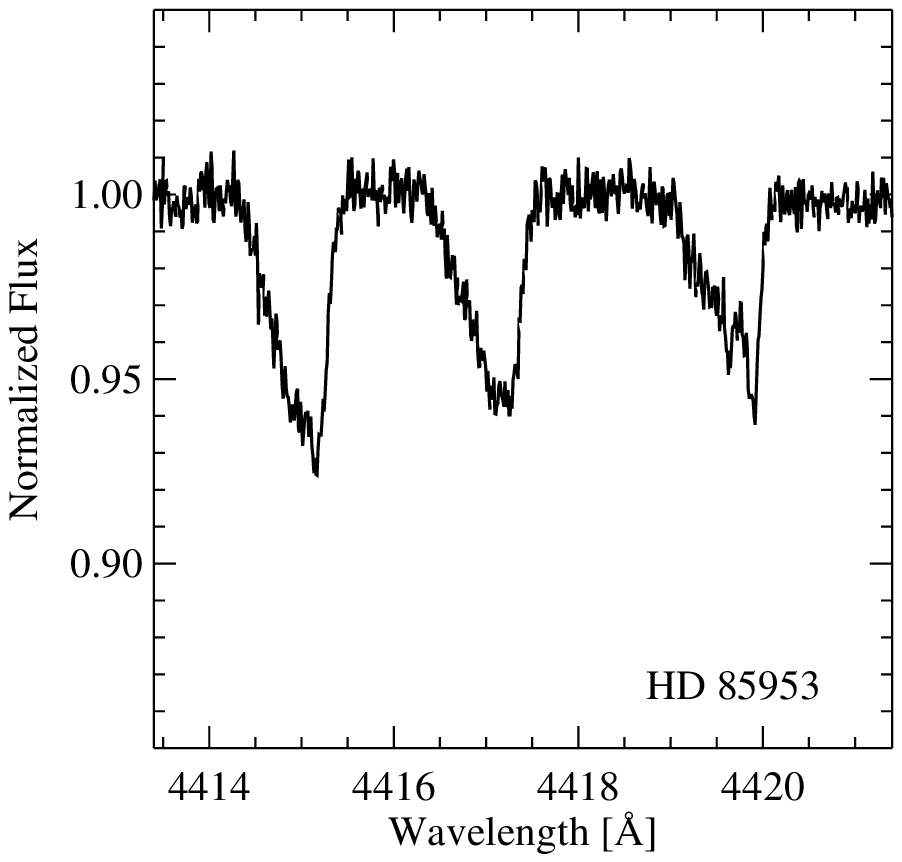}
\vspace{-5mm}
\caption[]{
Spectral variability as seen in HARPS Stokes~$I$ spectra.
{\em Left:} HD\,74195,
{\em middle:} HD\,74560,
{\em right:} HD\,85953.
}
\label{fig:schoeller_spb_var}
\end{center}
\end{figure}


The pulsation amplitudes
for the three studied pulsating stars
range from 4.5 to 25\,mmag.
Our study of correlations between the strength
of magnetic fields and pulsational characteristics
(Hubrig et al.\ 2009a \cite{schoeller_Hubrig2009a}) indicates
that it is possible that stronger magnetic fields appear in stars
with lower pulsating frequencies and smaller pulsating amplitudes.
Spectra for all three sources can be found in Fig.~\ref{fig:schoeller_spb_var}.
Spectral variability is evident for the two objects with more than
one observation.


\begin{figure}[!t]
\begin{center}
 \includegraphics[width=0.45\textwidth]{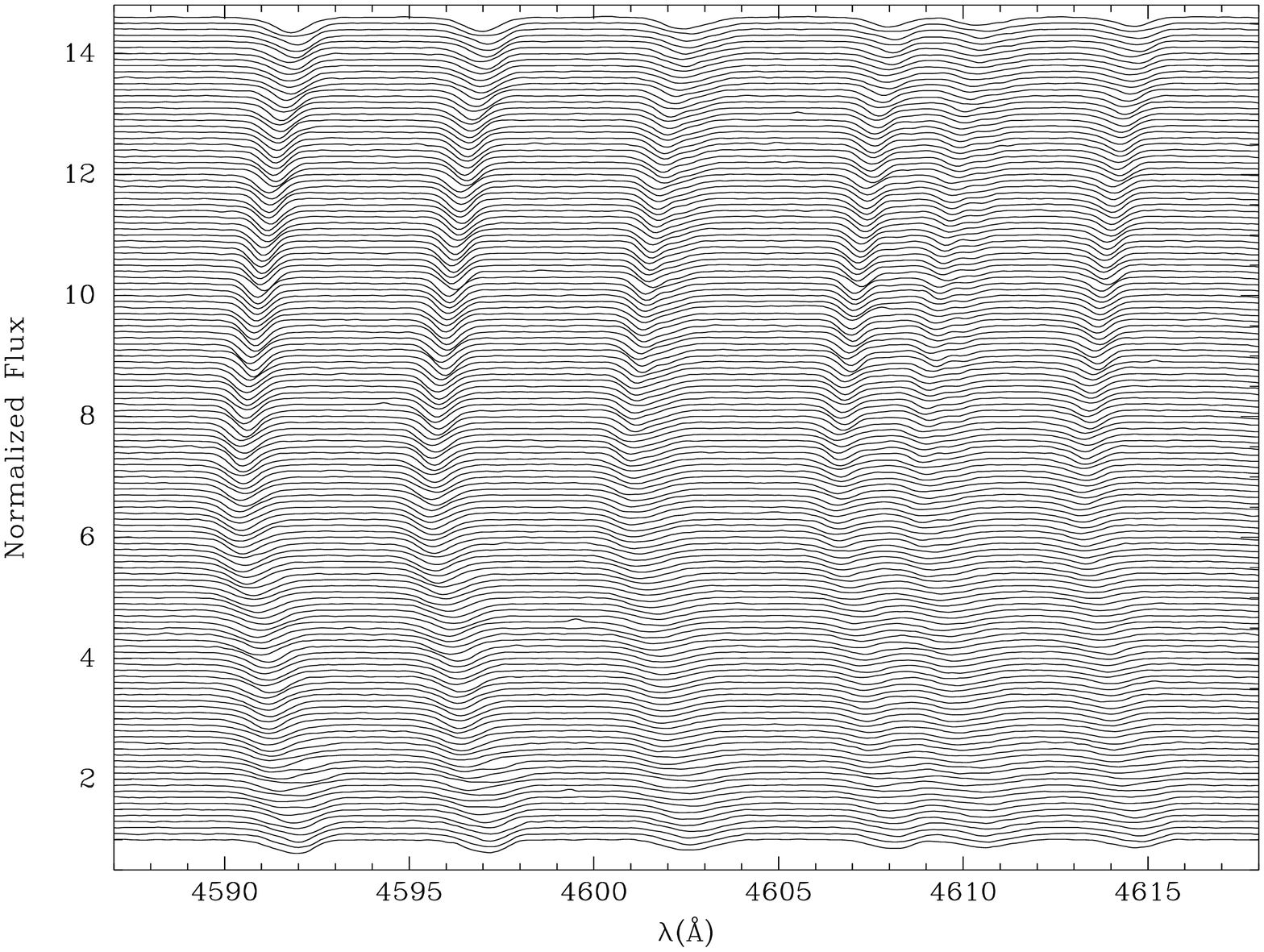}
 \includegraphics[width=0.45\textwidth]{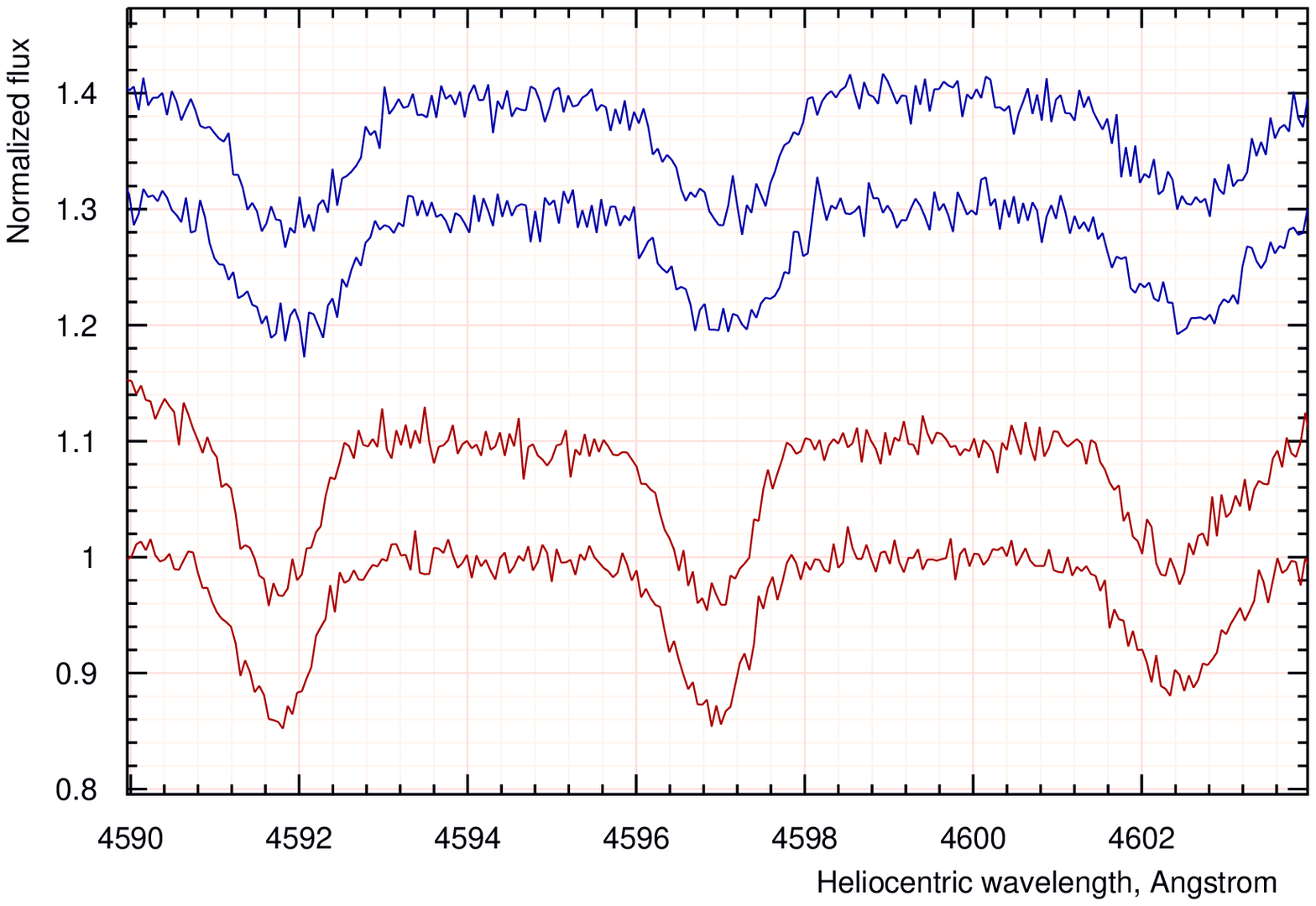}
\vspace{-5mm}
\caption[]{
{\em Left:}
Time series of FEROS spectra for V1449\,Aql showing pulsational line profile variability
in the spectral region 4590--4615\,\AA{}.
The pulsation phase zero is at the bottom.
{\em Right:}
Variability of the output spectra in two SOFIN sub-exposures taken with
the quarter-wave plate angles separated by $90^\circ$
taken around HJD\,2455398.530.
The lower two spectra, $(I+V)_0$ and $(I-V)_0$,
correspond to the first sub-exposure,
while the upper spectra, $(I-V)_{90}$ and $(I+V)_{90}$,
correspond to the second sub-exposure.
The strong effect of pulsations on the line profile shapes
and the line positions is clearly visible
between the spectra of the first sub-exposure
with a duration of 20\,min and the spectra of the second sub-exposure
with the same duration.
}
\label{fig:schoeller_v1449aql_var}
\end{center}
\end{figure}

From FEROS time series, one can find
line profile variability for V1449\,Aql with a pulsating frequency of
$f_{\rm puls} = 5.487$\,d$^{-1}$.
The variability in the spectra of V1449\,Aql and the impact of pulsations
on the polarimetric spectra can be seen in Fig.~\ref{fig:schoeller_v1449aql_var}.
Neglecting pulsations in the analysis of spectropolarimetric data will
lead to non-detections of magnetic fields in these stars
(Hubrig et al.\ 2011b \cite{schoeller_Hubrig2011b}).

\subsection{Be stars}
\label{subsect:schoeller_mf_be}

Rapidly rotating Be stars lose mass and initially accumulate it
in a rotating circumstellar disk.
Much of the mass loss is in the form of outbursts
and thus additional mechanisms such as the beating of nonradial pulsation modes
or magnetic flares must be at work.
Indirect evidence for the presence of a magnetic field are
variations of X-ray emission and
transient features in absorption line profiles.
Angular momentum transfer to a circumstellar disk,
channeling stellar wind matter, and accumulation of material
in an equatorial disk are more easily explained if magnetic fields can be invoked.
15 Be stars have been measured with the hydrogen polarimeter by Barker et al.\ (1985 \cite{schoeller_Barker1985})
using H$\beta$ - no detection was achieved. 
One Be star with a reported magnetic field,
$\omega$\,Ori (Neiner et al.\ 2003 \cite{schoeller_Neiner2003}),
was not confirmed as magnetic by recent observations.

\begin{figure}[!t]
\begin{center}
 \includegraphics[width=0.45\textwidth]{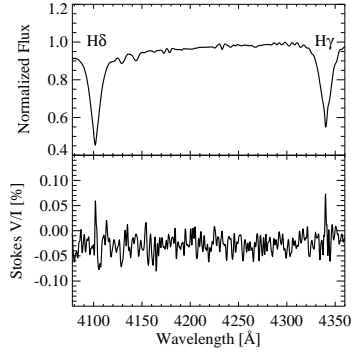}
\vspace{-5mm}
\caption[]{
Stokes~$I$ and Stokes~$V$ spectra of the Be star $o$\,Aqr ($\left<B_{\rm z}\right> = 98\pm31$\,G)
in the region including the H$\delta$ and H$\gamma$ lines.
}
\label{fig:schoeller_oaqr}
\end{center}
\end{figure}



\begin{figure}[!t]
\begin{center}
 \includegraphics[width=0.45\textwidth]{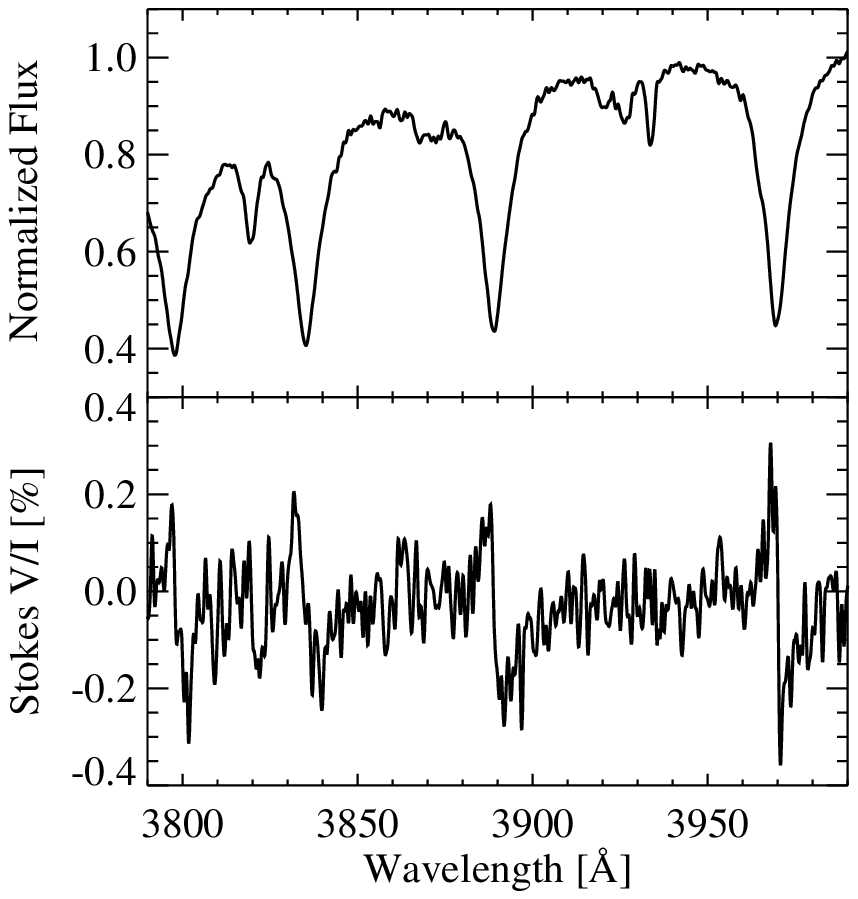}
 \includegraphics[width=0.45\textwidth]{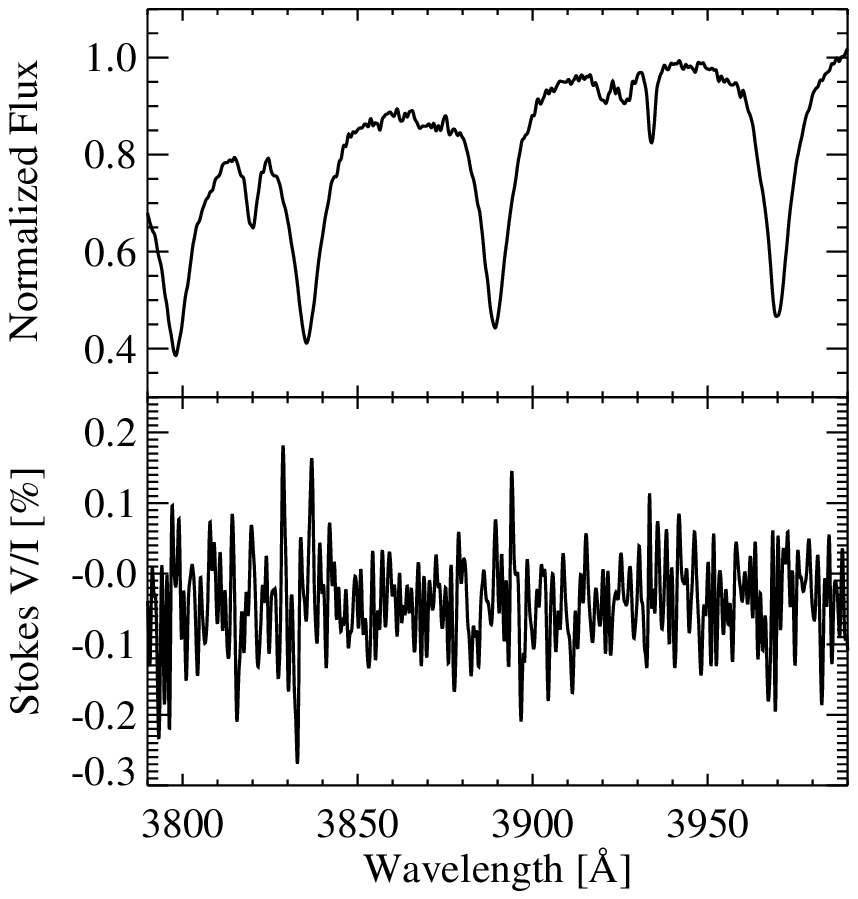}
\vspace{-5mm}
\caption[]{
{\em Left:} Stokes~$I$ and Stokes~$V$ spectra in
the blue spectral region around high number Balmer lines
of the He peculiar member NGC\,3766-170 of the young open cluster NGC\,3766
with the magnetic field
$\left<B_{\rm z}\right> = 1559\pm38$\,G, measured on hydrogen lines.
{\em Right:} Stokes~$I$ and Stokes~$V$ spectra
around high number Balmer lines for the candidate Be star NGC\,3766-45, with a magnetic field
$\left<B_{\rm z}\right> = -194\pm62$\,G measured on hydrogen lines.
}
\label{fig:schoeller_ngc3766}
\end{center}
\end{figure}

A sample of Be stars in the field and in the cluster NGC\,3766 (14.5--25\,Myr old)
was observed in 2006-2008 with FORS\,1.
A few Be stars show weak magnetic fields with the strongest field detected
in HD\,62367 ($\left<B_{\rm z}\right> = 117\pm38$\,G, $m_V = 7.1$).
Usually, the detected magnetic fields are below 100\,G
(see Figs.~\ref{fig:schoeller_oaqr} and \ref{fig:schoeller_ngc3766}).
The cluster NGC\,3766 appears to be extremely interesting,
where we find evidence for the presence of a magnetic field
in seven early-B type stars (among them three Be stars)
out of the observed 14 cluster members (Hubrig et al.\ 2009b \cite{schoeller_Hubrig2009b}).


\begin{figure}[!t]
\begin{center}
 \includegraphics[width=0.45\textwidth]{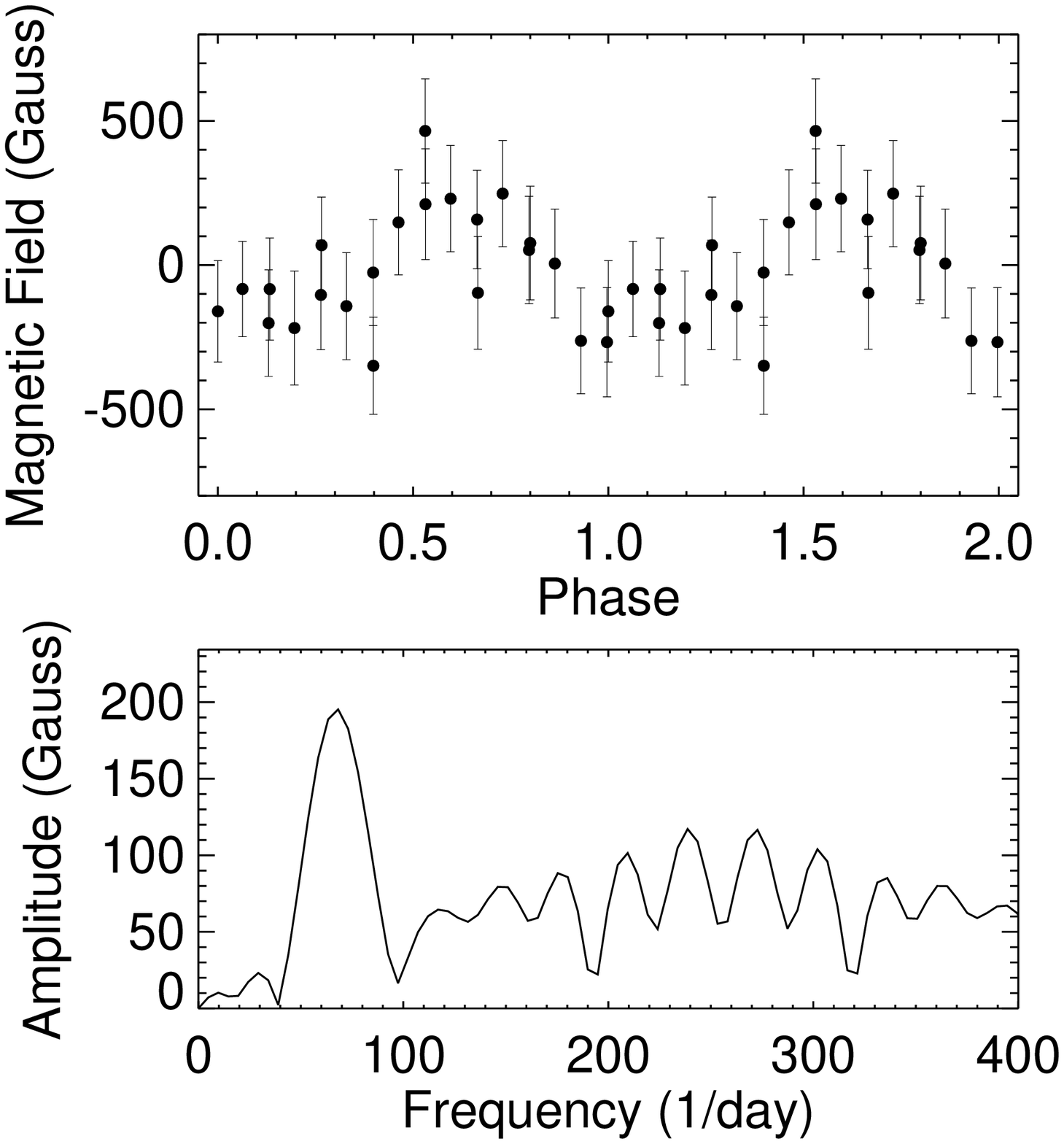}
 \includegraphics[width=0.45\textwidth]{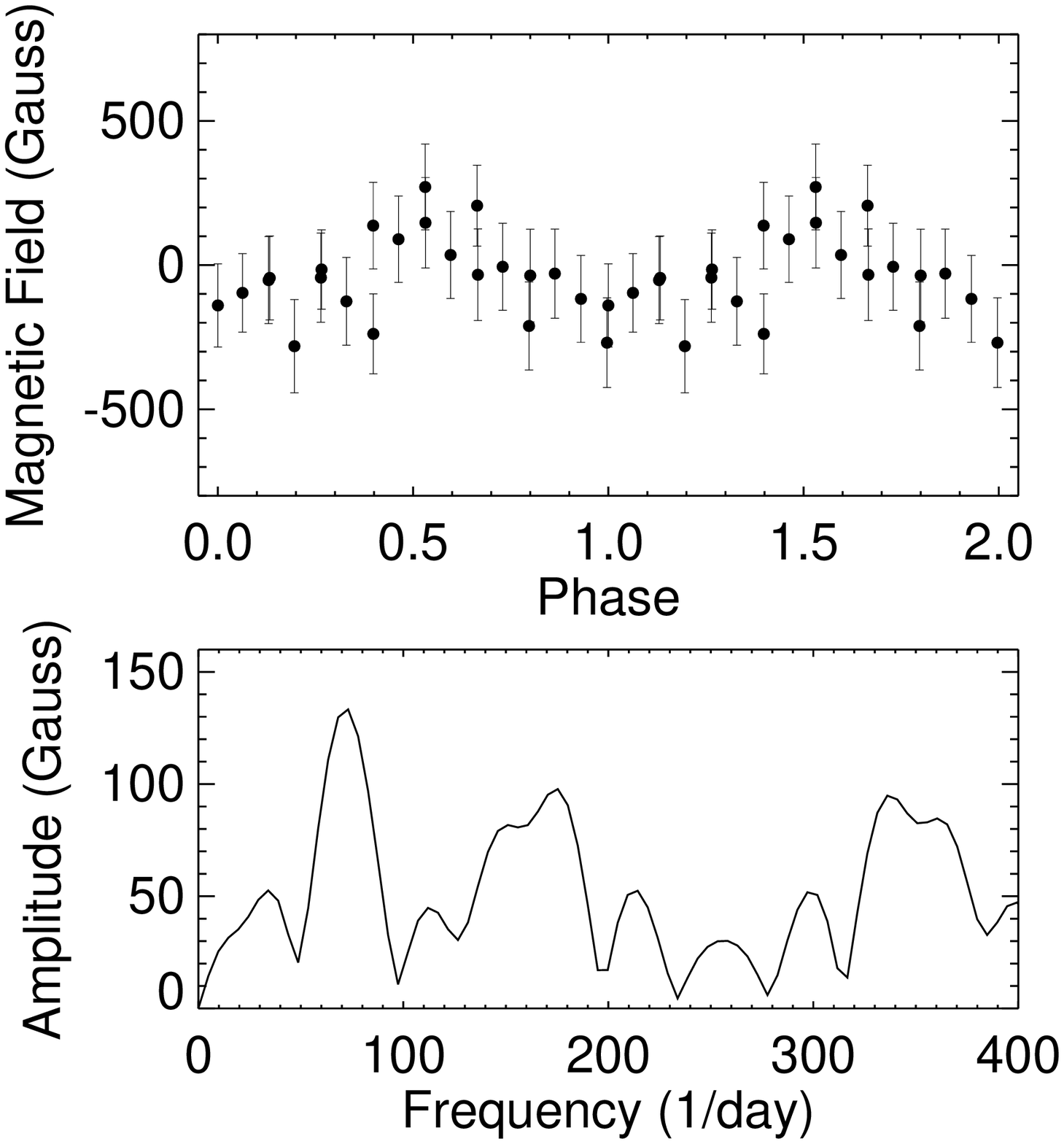}
\caption[]{
Phase diagram and amplitude spectrum for the magnetic field measurements of
$\lambda$\,Eri in 2006 August using hydrogen lines (left) and
all lines (right).

}
\label{fig:schoeller_lameri_phase_ampl}
\end{center}
\end{figure}

\begin{figure}[!t]
\begin{center}
 \includegraphics[width=0.45\textwidth]{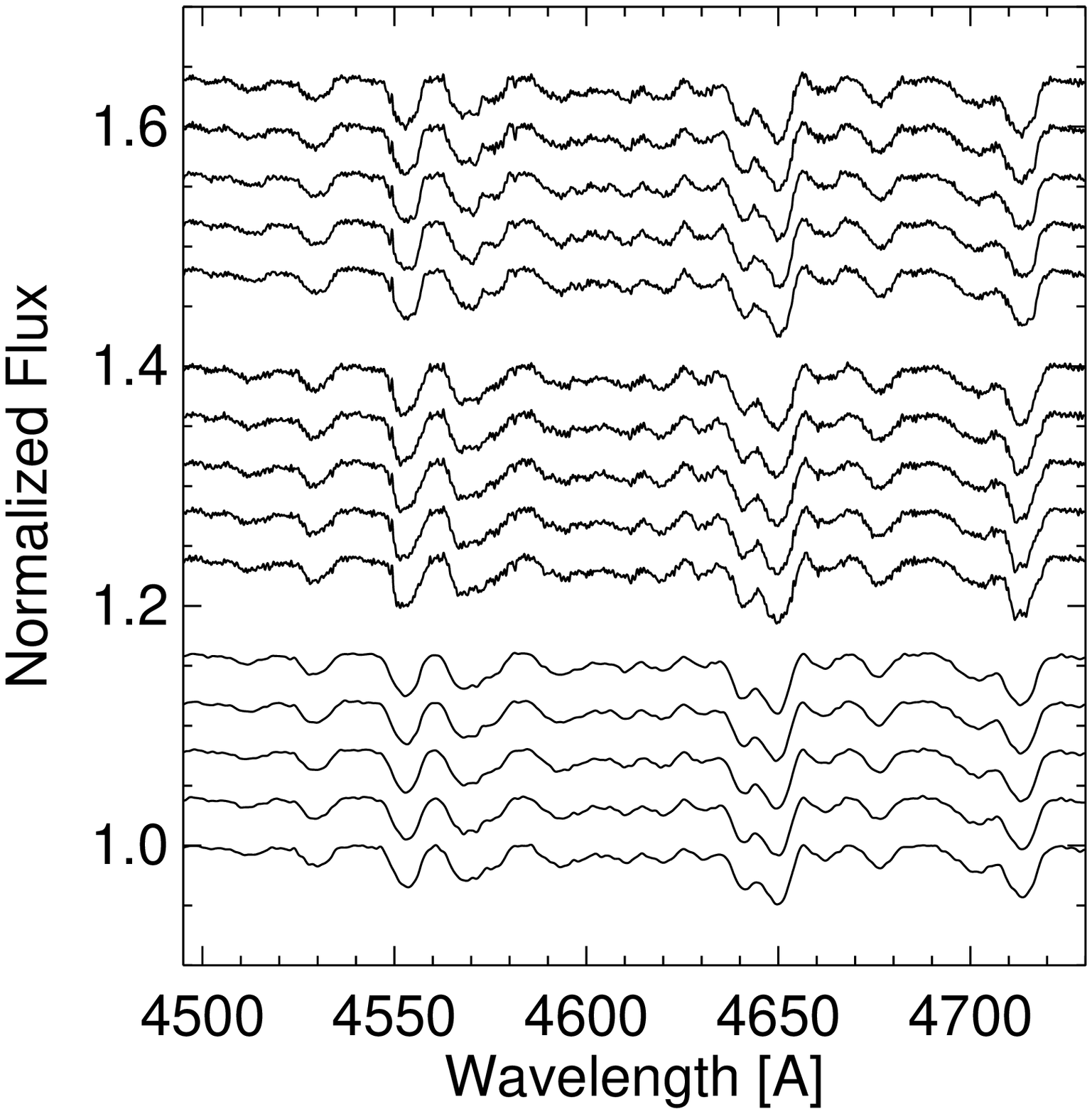}
\caption[]{
Spectrum variability of $\lambda$\,Eri on three different nights:
on 2006 August 8 (bottom),
2007 November 27 (middle), and
2007 November 28 (top).
}
\label{fig:schoeller_lameri_var}
\end{center}
\end{figure}

For nine early type Be stars, we obtained time-resolved magnetic field measurements
over $\sim$one hour  (up to 30 measurements) with FORS\,1 at the VLT.
For $\lambda$\,Eri, we were able to detect a period of $P=21.1$\,min
in the magnetic field measurements (see Fig.~\ref{fig:schoeller_lameri_phase_ampl}).
The spectral line profiles of $\lambda$\,Eri exhibit short-time periodic variability
(see Fig.~\ref{fig:schoeller_lameri_var})
due to non-radial pulsations with a period of 0.7\,d (Kambe et al.\ 1993 \cite{schoeller_Kambe1993}).
Furthermore, Smith (1994 \cite{schoeller_Smith1994}) detected dimples with a duration of 2--4\,h.
Do we see strong local magnetic fields?


Apart from $\lambda$\,Eri, four other stars showed indications of magnetic cyclic variability
on the scales of tens of minutes (Hubrig et al.\ 2009b \cite{schoeller_Hubrig2009b}).
A similar magnetic field periodicity ($P = 8.8$\,min) was detected
for the B0 star $\theta$\,Car (Hubrig et al.\ 2008 \cite{schoeller_Hubrig2008}).
These stars are good candidates for future time-resolved magnetic field observations
with high-resolution spectropolarimeters.

\section{Conclusions}
\label{sect:schoeller_concl}

Our magnetic field measurements using various spectropolarimetric instruments
have revealed the presence of magnetic fields in a number of different B-type stars,
including SPB, $\beta$\,Cep, and Be stars. 
New high-resolution spectropolarimetric observations with HARPS
support the magnetic nature of the studied stars. 
Future observations are urgently needed to determine
the role of magnetic fields in these objects.


%
%
%
%